\documentclass[a4paper]{jpconf}
\usepackage{graphicx}
\usepackage{psfrag}
\newcommand{\p}{\vec{p}}
\begin{document}
\title{New ways to access the transverse spin content of the nucleon}

\author{M. El Beiyad$^{1,2}$, B. Pire$^1$, L. Szymanowski$^3$ and S. Wallon$^{2,4}$}

\address{$^1$ CPHT, \'Ecole polytechnique, CNRS, 91128 Palaiseau, France}

\address{$^2$ LPT, Universit\'e d'Orsay, CNRS, 91404 Orsay, France}

\address{$^3$ Soltan Institute for Nuclear Studies, Warsaw, Poland }

\address{$^4$ UPMC, Univ. Paris 06, Facult\'e de physique, 75252 Paris, France }

\ead{pire@cpht.polytechnique.fr}

\begin{abstract}
We first describe a new way to access the chiral odd transversity parton distribution in the proton through the photoproduction of lepton pairs. The basic ingredient is the interference of the usual Bethe-Heitler or Drell-Yan amplitudes with the amplitude of a process, where the photon couples to quarks through its chiral-odd distribution amplitude, which is normalized to the magnetic susceptibility of the QCD vacuum.
We also show how the chiral-odd transversity generalized parton distributions (GPDs) of the nucleon can be accessed experimentally through the exclusive electro - or photoproduction process of a meson pair with a large invariant mass and when the final nucleon has a small transverse momentum. We calculate perturbatively the scattering amplitude at leading order, both in the high energy domain which may be accessed in electron-ion colliders and in the medium energy range. Estimated rates are encouraging.
\end{abstract}

\section{Introduction}
Transversity quark distributions in the nucleon remain among the most unknown leading twist hadronic observables. This is mostly due to their chiral odd character which enforces their decoupling in most hard amplitudes. After the pioneering studies \cite{tra}, much work \cite{Barone} has been devoted to the exploration of many channels but experimental difficulties have challenged the most promising ones. The recent focus on transversity dependent observables in single inclusive deep inelastic scattering propose to assume the factorization of transverse momentum dependent parton distributions (TMDs) and fragmentation functions. This allowed recently some first extraction of chiral odd quantities,  demonstrating, although  in a weak sense, their non-zero value. Here we restrict on the use of leading twist factorizable quantities, such as integrated parton distributions, distribution amplitudes or generalized parton distributions.

\section{Photoproduction of lepton pairs}
A new way to access the chiral odd transversity parton distribution in the proton emerges thanks to the observation that the photon twist 2 distribution amplitude (DA) \cite{Braun}  is chiral-odd.  This latter object is normalized to the magnetic susceptibility of the QCD vacuum. It is defined as
\begin{eqnarray}\label{def3:phi}
\langle 0 |\bar q(0) \sigma_{\alpha\beta} q(x) 
   | \gamma^{(\lambda)}(k)\rangle 
=       
 i \,e_q\, \chi\, \langle \bar q q \rangle
 \left( \epsilon^{(\lambda)}_\alpha k_\beta-  \epsilon^{(\lambda)}_\beta k_\alpha\right)  
 \int\limits_0^1 \!dz\, e^{-iz(kx)}\, \phi_\gamma(z)\,, \nonumber
\label{phigamma}
\end{eqnarray}    
where the normalization is chosen as $\int dz\,\phi_\gamma(z) =1$, 
and $z$ stands for the momentum fraction carried by the quark. The product of the quark condensate and of the magnetic susceptibility of the QCD vacuum
$\chi\, \langle \bar q q \rangle$ has been estimated  to be of the order of 50 MeV.  

The basic ingredient \cite{PS} is the interference of the usual Bethe Heitler or Drell-Yan amplitudes with the amplitude of a process, where the photon couples to quarks through this chiral-odd DA. We thus  consider the  process ($s_T$ is the transverse polarization vector of the nucleon):
\begin{equation}
\label{process}
\gamma(k,\epsilon) N (r,s_T)\to l^-(p)  l^+(p') X\,,
\end{equation}
with $q= p+p'$ in the kinematical region where $Q^2=q^2$ is large and the transverse component $ |\vec Q_\perp |$ 
of $q$ is of the same order as $Q$.
Such a process  occurs either through a Bethe-Heitler amplitude (Fig. 1a) where the initial photon 
couples to a final lepton, or through Drell-Yan type amplitudes (Fig. 1b) where the final leptons originate from 
a virtual photon. Among these Drell-Yan processes, one must distinguish the cases where the real photon couples 
directly (through the QED coupling) to quarks or through its quark content, {\em i.e.} the photon structure function. 
Gluon radiation at any order in the strong coupling $\alpha_s$, does not  introduce any chiral-odd 
quantity if one neglects quark masses.

We next consider the  contributions where the photon couples to the strong 
interacting particles through its twist-2 distribution amplitude (Fig. 1c and 1d). 
\begin{figure}
\includegraphics[width=3.8cm]{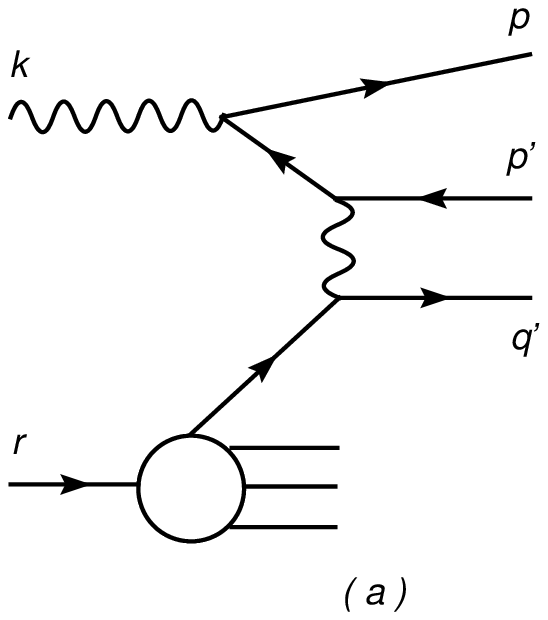}
\includegraphics[width=3.8cm]{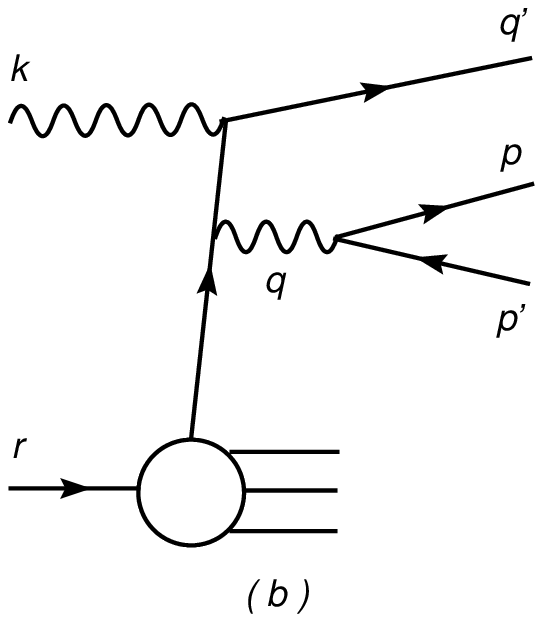}
\includegraphics[width=3.8cm]{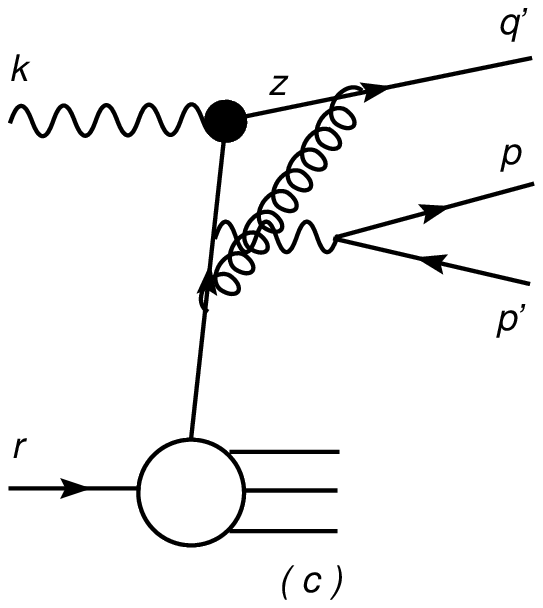}
\includegraphics[width=3.8cm]{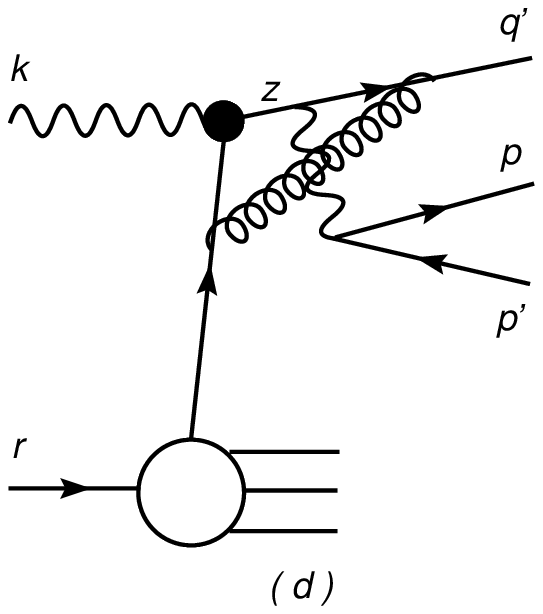}
\caption{Some amplitudes contributing to lepton pair photoproduction. (a) : The Bethe-Heitler process. (b) : The Drell-Yan process with the photon pointlike coupling. (c) -(d) : The Drell-Yan process with the photon Distribution Amplitude. }
\label{fig1}
\end{figure}
One can easily see by inspection that this  is the only way to get at the level of twist 2 (and with vanishing quark masses) 
a contribution to  nucleon transversity dependent observables.
We call this amplitude ${\cal A}_\phi$ :
\begin{eqnarray}\label{AmpC}
{\cal A}_\phi (\gamma q \to l \bar l q) \! = \! 2i \frac{C_F}{4N_c} e_q^2 e 4\pi\alpha_s \frac{ \chi\, \langle \bar q q \rangle }
{Q^2} \! \int \! dz \phi_\gamma (z)  \bar u(q') [ \frac{A_1}{x\bar z s (t_1+i\epsilon)} +\frac{A_2}{ z u (t_2+i\epsilon)} ]u(r) \bar u(p)\gamma^\mu v(p') \,,\nonumber
\end{eqnarray} 
with  $t_1= (zk-q)^2$ and $t_2= (\bar z k -q)^2$.
$
A_1 = x \,\hat r\,\hat \epsilon\,\hat k \,\gamma^\mu + \gamma^\mu \,\hat k \,\hat \epsilon \,\hat q $ and
$A_2 = \hat \epsilon\,\hat q\, \gamma^\mu \,\hat k + \hat k \,\gamma^\mu \,\hat q \,\hat \epsilon$
 do not depend on the light-cone fraction $z$.
${\cal A}_\phi$ develops an absorptive part  proportional to
\begin{eqnarray}
\label{abs}
\int dz \phi_\gamma (z) \bar u(q') [\frac{A_1}{x\bar z s}\delta (t_1) +\frac{A_2}{ z u} \delta (t_2)]u(r) \bar u(p)\gamma^\mu v(p')\,. \nonumber
\end{eqnarray}
This allows to perform the $z-$integration, the result of which, after using the $z-\bar z$ symmetry of the distribution amplitude, yields an absorptive part of the  amplitude ${\cal A}_\phi$ proportional to 
$\phi_\gamma (\frac{\alpha Q^2}{ Q^2+\vec Q_\perp ^2})$. This  absorptive part, which may be measured in single spin asymmetries, as discussed below, thus scans the photon chiral-odd distribution amplitude.

The cross section for reaction (\ref{process}) can  be read as
\begin{eqnarray}\label{cs}
\frac {d\sigma}{d^4Q \,d\Omega}  =  \frac {d\sigma_{BH}}{d^4Q\,d\Omega} +  \frac {d\sigma_{DY}}{d^4Q\,d\Omega} +   \frac {d\sigma_{\phi}}{d^4Q\,d\Omega} +  \frac {\Sigma d\sigma_{int}}{d^4Q\,d\Omega}\,,\nonumber
\end{eqnarray}
where $\Sigma d\sigma_{int}$ contains various interferences, while
the transversity dependent  differential cross section (we denote $\Delta_T \sigma = \sigma(s_T) - \sigma(-s_T)$) reads
\begin{eqnarray}\label{cst}
&&\frac {d\Delta_T \sigma}{d^4Q\,d\Omega}  = \frac {d\sigma_{\phi int}}{d^4Q\,d\Omega}  \,,
\end{eqnarray}
 where $d\sigma_{\phi int}$ contains only interferences between the amplitude ${\cal A}_\phi$ and the other amplitudes. Moreover, one may use the distinct charge conjugation property (with respect to the lepton part) of the Bethe Heitler amplitude to select the interference between ${\cal A}_\phi$ and the Bethe-Heitler amplitude :
\begin{eqnarray}\label{CAcst}
&&\frac {d\Delta_T \sigma (l^-) - d\Delta_T \sigma (l^+) }{d^4Q\,d\Omega}  = \frac {d\sigma_{\phi BH}}{d^4Q\,d\Omega}  \,.
\end{eqnarray}

The polarization average of $d\sigma_{\phi BH}$ reads :
\begin{eqnarray}
\frac {1}{2} \sum_\lambda d\sigma_{\phi BH} (\gamma(\lambda)p\to l^-l^+X) 
\! = \! \frac{(4\pi\alpha_{em})^3}{4 s } \, \frac{C_F 4 \pi\alpha_s}{2N_c}    \frac{\chi\, \langle \bar q q \rangle}{\vec Q_\perp ^2}\! \int \! dx \!\sum_q Q_l^3Q_q^3h_1^q(x) 2{\cal R}e({\cal I}_{\phi BH})\,dLIPS \,, \nonumber
\label{cspa}
\end{eqnarray}
with the usual Lorentz invariant phase space factor dLIPS  $= (2\pi)^4 \delta^4(P_{in}-P_{out}) \Pi \frac{d^3p_i}{2E_i(2\pi)^3} $, ($p_i=p,p',q'$)
and
\begin{eqnarray}
\label{1}
{2\cal R}e({\cal I}_{\phi BH}) &=& \phi_\gamma \left(\frac{\alpha Q^2}{ Q^2+\vec Q_\perp ^2}\right)\, \frac{32\pi \alpha^2 \bar \alpha}{xs (\bar \alpha Q^2+\vec Q_\perp ^2)^2} 
{ (Q^2 + \vec Q_\perp^2)} [\epsilon^{rks_TQ_T} A_1+ \epsilon^{rks_Tl_T} A_2]\,,\nonumber
\end{eqnarray}
where 
\begin{eqnarray}
\label{2}
&&A_1= \frac{2}{\alpha^2Q^2} [-2\vec l_\perp.\vec Q_\perp +\bar\alpha(\gamma-\bar \gamma)Q^2 ] 
+\frac{Q^2\,[2\vec l_\perp.\vec Q_\perp +(\gamma- \bar \gamma) \vec Q_\perp ^2]}
{ ( \gamma Q^2 -2\vec l_\perp.\vec Q_\perp +\bar  \gamma \vec Q_\perp ^2)(\bar \gamma Q^2 +2\vec l_\perp.\vec Q_\perp +  \gamma \vec Q_\perp ^2)} \,, 
 \nonumber
\end{eqnarray}
and 
\begin{eqnarray}
\label{3}
&&A_2=\frac{2}{\alpha^2Q^2} \,[\bar\alpha^2 Q^2 + \vec Q_\perp ^2] 
-\frac{\vec Q_\perp^2\,[Q^2+\vec Q_\perp ^2]}
{ ( \gamma Q^2 -2\vec l_\perp.\vec Q_\perp +\bar  \gamma \vec Q_\perp ^2)(\bar \gamma Q^2 +2\vec l_\perp.\vec Q_\perp +  \gamma \vec Q_\perp ^2)} \,. 
 \nonumber
\end{eqnarray}
This observable is proportional to the transversity nucleon distribution $h_1(x)$ at the position $x=\frac{Q^2}{\alpha s}+\frac{\vec Q_\perp ^2}{\alpha \bar\alpha s} $, which is fixed by the event kinematics. We are thus in a position to scan the transversity quark distribution. 

\section{Transversity GPDs}
Generalized parton distributions (GPDs) offer a new way to access the transversity dependent quark content of the nucleon. The factorization properties of exclusive amplitudes allows in principle to extract the four chiral-odd transversity GPDs ~\cite{defDiehl}, noted  $H_T$, $E_T$, $\tilde{H}_T$, $\tilde{E}_T$.
Their access is however quite challenging~\cite{DGP}  : one photon or one meson electroproduction leading twist amplitudes are insensitive to transversity GPDs. A possible way out is to consider higher twist contributions to these amplitudes \cite{liuti}, which however are beyond the factorization proofs and often plagued with end-point singularities. The strategy which we followed  in Ref.~\cite{IPST,eps} and in  Ref.~\cite{EPSSW}, is to study the leading twist contribution to exclusive processes where more mesons are present in the final state. A similar strategy has also been advocated recently in Ref.~\cite{kumano} to enlarge the number of processes which could be used to extract information on chiral-even GPDs.

\subsection{Diffractive photoproduction of two $\rho$ mesons}
 In the example developed in ~\cite{IPST,eps}, the process under study is the high energy photo (or electro) diffractive production of two vector mesons, the hard probe being the virtual "Pomeron" exchange (and the hard scale being the virtuality of this Pomeron), in analogy with the virtual photon exchange occuring in the deep inelastic electroproduction of a meson. 
 
 The chiral-odd light-cone DA for the transversely polarized meson vector $\rho^0_T$,  is defined, in leading twist 2, by the matrix element :
\begin{equation}
\langle 0|\bar{u}(0)\sigma^{\mu\nu}u(x)|\rho^0(p,\epsilon_\pm) \rangle = \frac{i}{\sqrt{2}}(\epsilon^\mu_{\pm}(p)\, p^\nu - \epsilon^\nu_{\pm}(p)\, p^\mu)f_\rho^\bot\int_0^1du\ e^{-iup\cdot x}\ \phi_\bot(u),
\label{defDArho}
\end{equation}
where $\epsilon^\mu_{\pm}(p_\rho)$ is the $\rho$-meson transverse polarization and with $f_\rho^\bot$ = 160 MeV.
We consider the specific process :
\begin{equation}
\label{2mesongen}
\gamma^*_L (q)\;\; N (p_2) \to  \rho_{L}^0(q_\rho)\;\; \rho^+_{T}(p_\rho)\;
N'(p_{2'})\;,
\end{equation}
 of scattering of a  virtual  photon on a
nucleon $N$, which leads via two gluon exchange to the production
of  two vector mesons separated by a large
rapidity gap  and the scattered nucleon $N'$. We choose a charged vector meson $\rho^+$ since it requires
 quark antiquark exchange in the $t-$channel with the nucleon line.
We consider the kinematical region where the rapidity gap 
between $\rho^+$ and
$N'$ is much smaller than the one between
$\rho^0$ and $\rho^+$, that is the energy of the system 
($\rho^+\; -\; N'$) is smaller
than the energy of the system ($\rho^0 -\; \rho^+$) but still large enough to
justify our approach  (in particular much larger than baryonic resonance masses).

We have shown that in  such kinematical
circumstance, the Born term for this process is calculable consistently within
the collinear factorization method. The final result is represented as an
integral (over  the longitudinal momentum fractions of the quarks)  of
the product of two amplitudes: the first one describing
the transition $\gamma^* \to \rho_L^0$ via two gluon exchange and
the second one  describing the subprocess
${\cal P}\;N\;\to \;\rho^+\;N'$ 
which is
closely related to the electroproduction process 
$\gamma^*\,N \to\rho^+\,N'$
where  collinear factorization theorems allow to separate  the long distance dynamics  expressed
through the GPDs from a perturbatively calculable coefficient function. The case of transversally
 polarized vector meson $\rho_{T}^+$  involves the chiral-odd GPD. The hard scale
appearing in this process  is supplied by the relatively large  momentum transfer
$p^2$ in the two gluon channel, i.e. by the virtuality of the Pomeron.

We have shown that the collinear factorization holds
at least in the Born approximation. 
The resulting scattering amplitude ${\cal M}^{\gamma^*\,p\,\to \rho_L^0\, \rho^+_T\,n}$ has a very 
compact form  : 

\begin{eqnarray}
\label{CON}
&&{\cal M}^{\gamma^{(*)}_{L/T} \,p\,\to \rho_L^0\, \rho^+_T\,n}
\nonumber \\
&&= -\,\sin \theta \;16\pi^2 s \alpha_s f_\rho^T \xi \sqrt\frac{1-\xi}{1+\xi}
\frac{C_F}{N\,(\p^{\;2})^2}
\times\int\limits_0^1
\frac{\;du\;\phi_\perp(u)}{ \,u^2 \bar u^2 }
 J^{\gamma^{(*)}_{L/T} \to \rho^0_L}(u\p,\bar u\p)
 H^{u\,d}_T(\xi(2u-1),\xi,0),
 \nonumber \\
 && H^{u\,d}(\xi(2u-1),\xi,0)=\left[ H_{T}^u(\xi (2u-1),\xi,0)- H_T^{d}(\xi (2u-1),\xi,0)\right]
\end{eqnarray}
where  $\theta$ is the angle between the transverse polarization vector of
the target
$\vec{n}$ and the polarization vector
$\vec{\epsilon}_T$ of the produced $\rho^+_T-$meson. The 
flavour non-diagonal transversity GPD $H^{u\,d}_T$ 
between an initial proton and a final neutron 
is the difference  of the flavour diagonal transversity GPDs  $H^{u}_T$, 
$H^{d}_T$ of quarks $u$ and $d$ inside a proton. We assumed also 
that the Mandelstam variable $-t= -(p_2-p_{2'})^2$ takes its minimal value 
which means that only the transversity GPD $H_T$ contributes to the process
(\ref{2mesongen}).

\begin{figure}
\includegraphics[width=7.8cm]{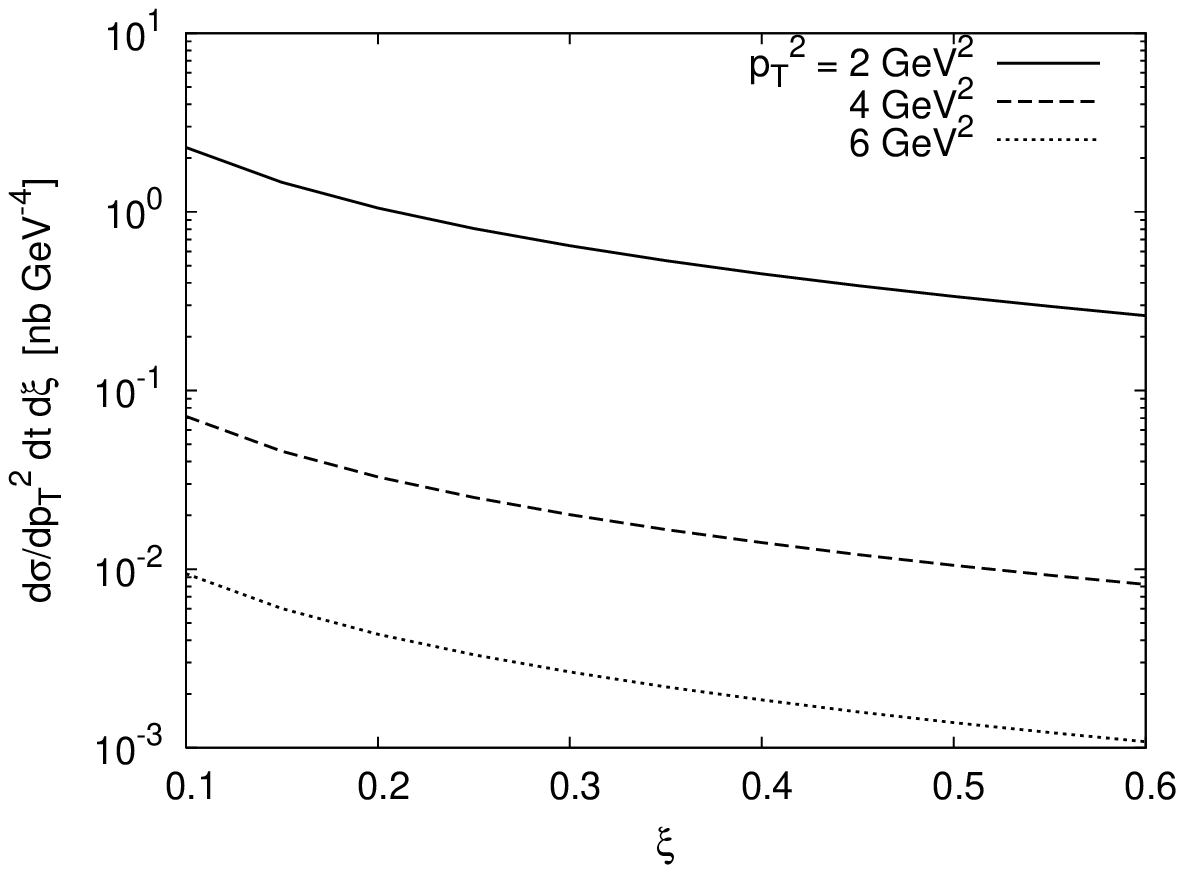}
\includegraphics[width=7.8cm]{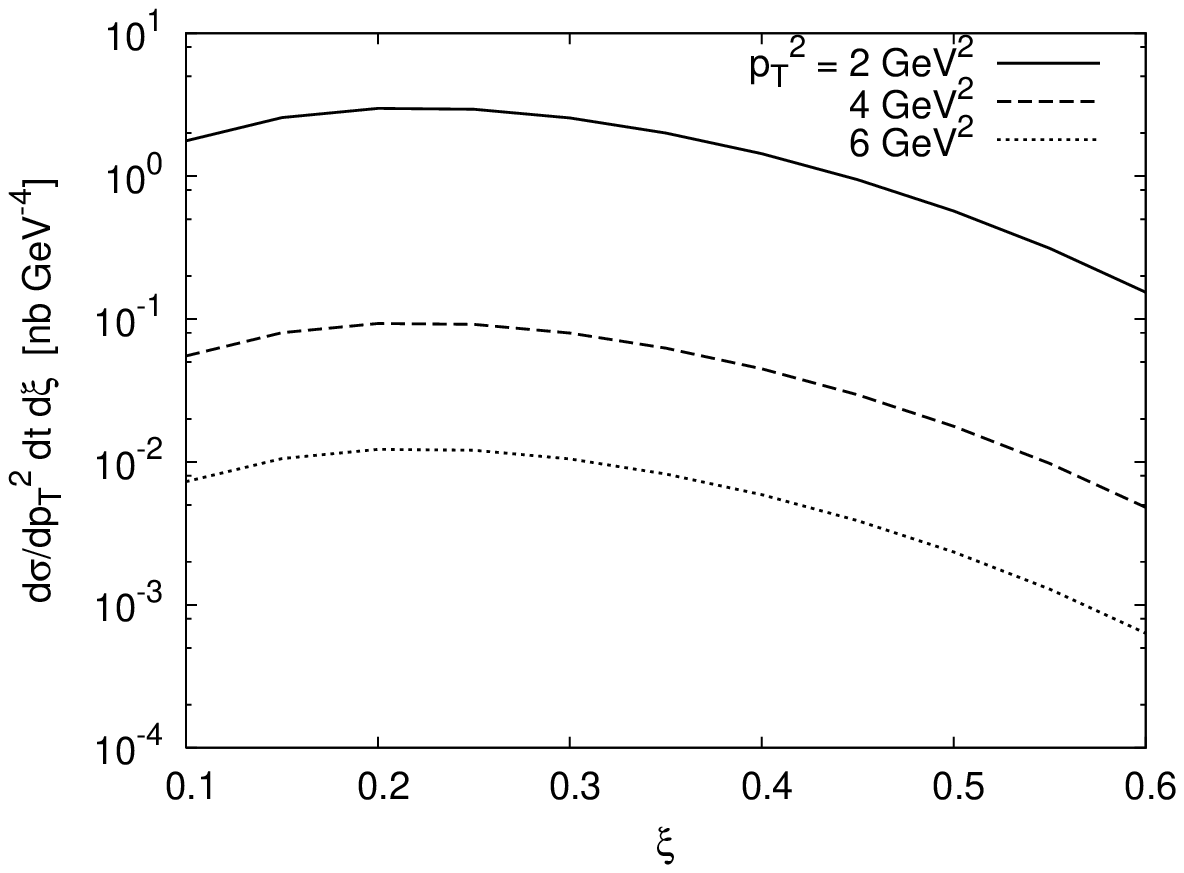}
\caption{The differential cross section for the photoproduction of transversely polarized 
$\rho^0$ and $\rho^+$ (left figure) and of longitudinally polarized 
$\rho^0$ and $\rho^+$ (right figure) as a function of $\xi$ for $p_T^2 = 2,\,4,\,$ and $6$ GeV$^2$.}
\label{fig2}
\end{figure}

$J^{\gamma^*_L \to \rho^0_L}$ is the  impact factor 
\begin{equation}
\label{ifgamma}
J^{\gamma^*_L \to \rho^0_L}(\vec{k}_1,\vec{k}_2)=  -   f_\rho \frac{e
\alpha_s 2\pi
 Q}{N_c\sqrt{2}} \int\limits_0^1 dz\;z\bar z \phi_{||}(z)P(\vec{k}_1,\vec{k}_2)\;, \nonumber
\end{equation}
with $P(\vec{k}_1,\vec{k}_2=\vec{p}-\vec{k}_1)$ written as :
\begin{equation}
\label{P}
\frac{1}{z^2\vec{p}^{\;2}+m_q^2 +Q^2z\bar z} +
\frac{1}{{\bar z}^2\vec{p}^{\;2}+m_q^2 +Q^2z\bar z} -\frac{1}{(\vec{k}_1-z\vec{p}\,)^2+m_q^2 +Q^2z\bar z} -
\frac{1}{(\vec{k}_1-\bar z\vec{p}\,)^2+m_q^2 +Q^2z\bar z}. \nonumber
\end{equation}
One can write
similar formulas  for the impact factor describing the transition 
$\gamma^*_T \to \rho_L^0$. The scattering amplitude (\ref{CON}) receives a contribution only from the 
ERBL region. Thus one needs a model for the $H_T^{u,\,d}$ GPDs in this region. For this aim we 
generalized the model of nucleon tensor charge proposed in 
the Ref. \cite{GG}
to the non-forward kinematics.
 In order to make our preditions more reliable we compared our estimates for the 
process (\ref{2mesongen}) which involves the transversity GPD with a similar proces 
$\gamma^*\,p\,\to \rho_L^0\, \rho^+_L\,n$ of production of two longitudinally polarized 
$\rho$ mesons, which is sensitive to the usual chiral-even GPDs. We modellized the chiral-even GPDs 
 using method based on 
the double distribution. The comparison of the differential cross sections for unpolarized beam and target
\begin{equation}
\label{crosssec}
\frac{\mbox{d}\sigma}{\mbox{d}p_T^2\, \mbox{d}t\, \mbox{d}\xi} = 
\frac{1}{256\, \pi^3\, \xi(1-\xi)\, s^2} | \cal{M} |^{\mbox{2}}
\end{equation}
for both  processes
is shown in Fig.~\ref{fig2}. 
  Note that, contrarily to transversity PDFs, transversity GPDs enter the
formulae for cross sections even when considering unpolarized proton
target, provided one selects the polarization state of an outgoing meson.

\subsection{Photoproduction of $\pi \rho_T$ pair}

Diffractive physics requires high energies. In the realm of medium energies where higher luminosities are achieved, a QCD study based solely on the collinear factorization approach may be followed. We thus study \cite{EPSSW} the exclusive photoproduction of a transversely polarized vector meson and a pion on a  polarized or unpolarized proton target

\begin{equation}
\gamma(q) + N(p_1,\lambda) \rightarrow \pi(p_\pi) + \rho_T(p_\rho) + N'(p_2,\lambda')\,,
\label{process2}
\end{equation}
 in the kinematical regime of large invariant mass $M_{\pi\rho}$ of the final meson pair and small momentum transfer $t =(p_1-p_2)^2$ between the initial and the final nucleons.  The hard factorization scale is this invariant mass  $M_{\pi\rho}$. Roughly speaking, these kinematics mean a moderate to large, and approximately opposite, transverse momentum of each  meson.
 
 The amplitude gets contributions from each of the four twist 2 chiral-odd GPDs . However, all of them but $H_T$ are accompanied by kinematical factors which vanish at $\vec{\Delta}_t=0\,.$
The contribution  proportional to $H_T$ is thus dominant in the small $t$ domain which we are interested in and we restrict our study to this contribution, so that the whole $t-$dependence will come from the $t$-dependence of $H_T$.
The scattering amplitude of the process (\ref{process}) in the factorized form :
\begin{equation}
\label{AmplitudeFactorized}
\mathcal{A}(t,M^2_{\pi\rho},p_T)  =\frac{1}{\sqrt{2}} \int_{-1}^1dx\int_0^1dv\int_0^1dz\ (T^u(x,v,z)-T^d(x,v,z)) \, H^{ud}_T(x,\xi,t)\Phi_\pi(z)\Phi_\bot(v)\,,
\end{equation}
where $T^u$ and $T^d$ are the hard parts of the amplitude where the photon couples respectively to a $u$-quark  and to a $d$-quark. 
 
The amplitude of this process can be simplified as
\begin{eqnarray}
\label{ampl}
\mathcal{A} = (\vec{N}_t\cdot \vec{e}^{\,*}_\pm)(\vec{p}_t\cdot \vec{\epsilon}_{\gamma t}) A + (\vec{N}_t\cdot\vec{\epsilon}_{\gamma t})(\vec{p}_t\cdot\vec{e}^{\,*}_\pm) B+ (\vec{N}_t\cdot\vec{p}_t) (\vec{\epsilon}_{\gamma t}\cdot\vec{e}^{\,*}_\pm) C + (\vec{N}_t\cdot\vec{p}_t) (\vec{p}_t\cdot\vec{\epsilon}_{\gamma t}) (\vec{p}_t\cdot \vec{e}^{\,*}_\pm) D \nonumber
\end{eqnarray}
where $A$, $B$, $C$, $D$ are scalar functions of $s$, $\xi$, $\alpha$ and $M^2_{\pi\rho}$.
We model the transversity GPD $H_T^q(x,\xi,t)$ ($q=u,\ d$)  in terms of double distributions 
\begin{equation}
\label{DDdef}
H_T^q(x,\xi,t=0) = \int_\Omega d\beta\, d\alpha\ \delta(\beta+\xi\alpha-x)f_T^q(\beta,\alpha,t=0) \,,
\end{equation}
where $f_T^q$ is the quark transversity double distribution :
\begin{equation}
\label{DD}
f_T^q(\beta,\alpha,t=0) = \Pi(\beta,\alpha)\,\delta \, q(\beta)\Theta(\beta) - \Pi(-\beta,\alpha)\,\delta \bar{q}(-\beta)\,\Theta(-\beta)\,,
\end{equation}
where $ \Pi(\beta,\alpha)$ is a profile function and $\delta q$, $\delta \bar{q}$ are the quark and antiquark transversity parton distribution functions  parametrized in~\cite{Anselmino}. The resulting GPDs have the same order of magnitude but some
differences with other models  \cite{Pasq}. The $t$-dependence of these chiral-odd GPDs and its Fourier transform in terms of the transverse localization of quarks in the proton \cite{impact} are very interesting but completely unknown. We
 describe it in a simplistic way as:
\begin{equation}
\label{t-dep}
H^q_T(x,\xi,t) = H^q_T(x,\xi,t=0)\times  \frac{C^2}{(t  - C)^2},
\end{equation}
with  $C=.71~$GeV$^2$. We have no phenomenological control of this assumption, since  the tensor form factor of the nucleon  has never been measured.
\begin{figure}[!h]
$\begin{array}{cc}
\psfrag{fpi}{$\,\phi_\pi$}
\psfrag{fro}{$\,\phi_\rho$}
\psfrag{z}{\begin{small} $z$ \end{small}}
\psfrag{zb}{\raisebox{-.1cm}{ \begin{small}$\hspace{-.3cm}-\bar{z}$\end{small}} }
\psfrag{v}{\begin{small} $v$ \end{small}}
\psfrag{vb}{\raisebox{-.1cm}{ \begin{small}$\hspace{-.3cm}-\bar{v}$\end{small}} }
\psfrag{gamma}{$\,\gamma$}
\psfrag{pi}{$\,\pi^+$}
\psfrag{rho}{$\,\rho^0_T$}
\psfrag{N}{$N$}
\psfrag{Np}{$\,N'$}
\psfrag{H}{\hspace{-0.5cm} $H^{ud}_T(x,\xi,t_{min})$}
\psfrag{hard}{\hspace{-0.2cm} $H^{ud}_T(x,\xi,t_{min})$}
\psfrag{p1}{\begin{small}     $p_1$       \end{small}}
\psfrag{p2}{\begin{small} $p_2$ \end{small}}
\psfrag{p1p}{\hspace{-0.8cm}  \begin{small}  $p_1'=(x+\xi) p$  \end{small}}
\psfrag{p2p}{\hspace{-0.2cm} \begin{small}  $p_2'=(x-\xi) p$ \end{small}}
\psfrag{q}{\begin{small}     $q$       \end{small}}
\psfrag{ppi}{\begin{small} $p_\pi$\end{small}}
\psfrag{prho}{\begin{small} $p_\rho$\end{small}}
\includegraphics[width=7.0cm]{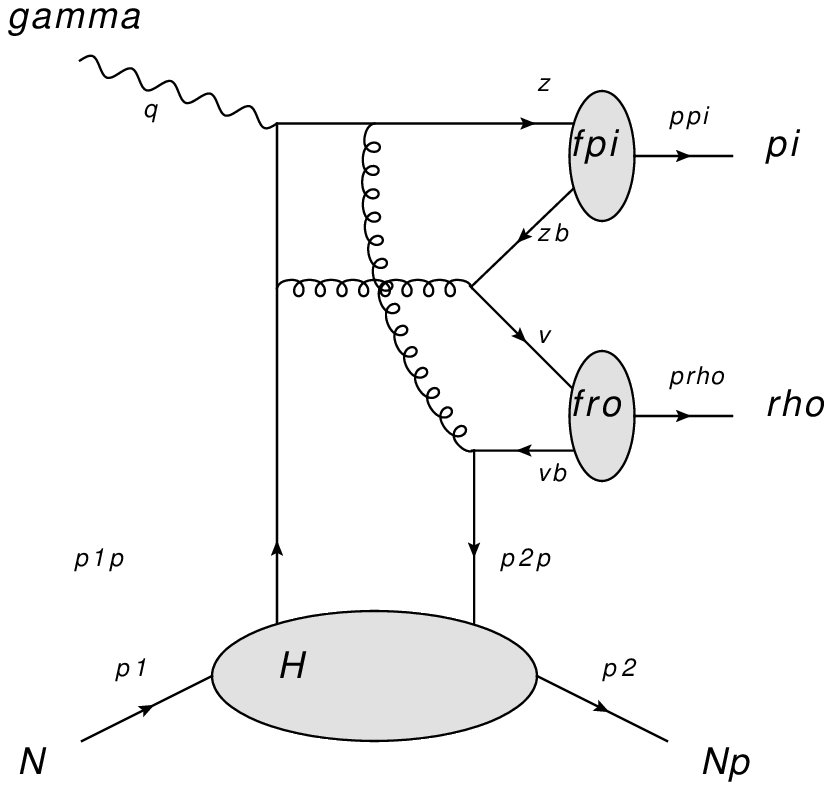}&\hspace{-0.5cm}
\psfrag{fpi}{$\,\phi_\pi$}
\psfrag{fro}{$\,\phi_\rho$}
\psfrag{z}{\begin{small} $z$ \end{small}}
\psfrag{zb}{\raisebox{-.1cm}{ \begin{small}$\hspace{-.3cm}-\bar{z}$\end{small}} }
\psfrag{v}{\raisebox{-.2cm}{\begin{small} $\hspace{-.2cm}-\bar{v}$ \end{small}}}
\psfrag{vb}{\raisebox{-.1cm}{ \begin{small}$v$\end{small}} }
\psfrag{gamma}{$\,\,\,\gamma$}
\psfrag{pi}{$\,\pi^+$}
\psfrag{rho}{$\,\rho^0_T$}
\psfrag{N}{$N$}
\psfrag{Np}{$\,N'$}
\psfrag{H}{\hspace{-0.5cm} $H^{ud}_T(x,\xi,t_{min})$}
\psfrag{p1}{\begin{small}     $p_1$       \end{small}}
\psfrag{p2}{\begin{small} $p_2$ \end{small}}
\psfrag{p1p}{\hspace{-0.8cm}  \begin{small}  $p_1'=(x+\xi) p$  \end{small}}
\psfrag{p2p}{\hspace{-0.2cm} \begin{small}  $p_2'=(x-\xi) p$ \end{small}}
\psfrag{q}{\begin{small}     $q$       \end{small}}
\psfrag{ppi}{\begin{small} $p_\pi$\end{small}}
\psfrag{prho}{\begin{small} $p_\rho$\end{small}}
\includegraphics[width=7.0cm]{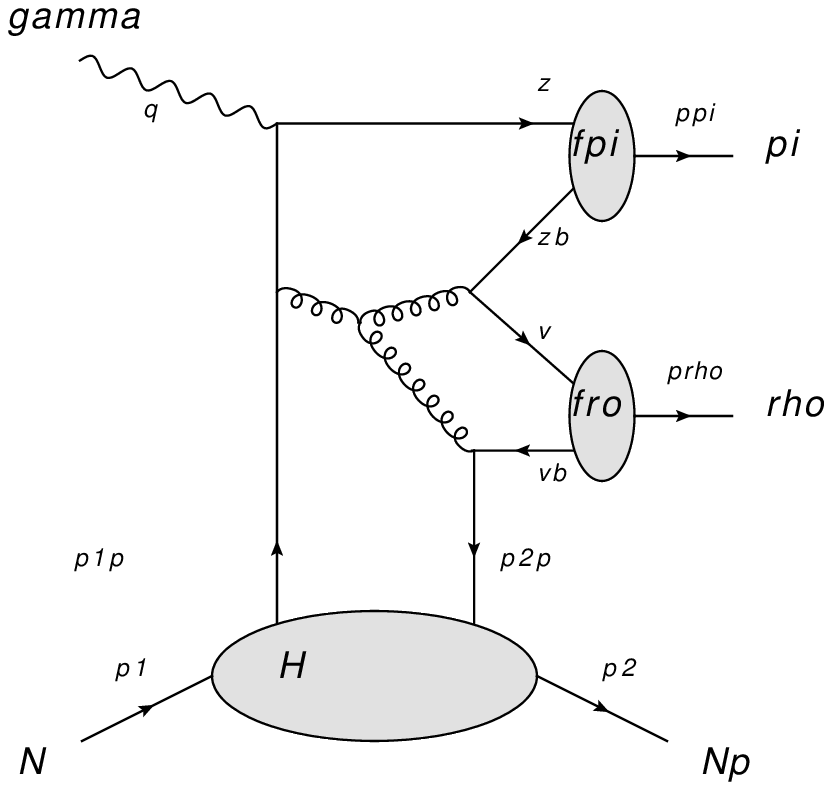}\\
\end{array}$
\caption{Two out of 62 representative diagrams contributing to the process (\ref{process}).}
\label{twodiagrams}
\end{figure}
The calculation of the hard part at the leading order in the strong 
coupling $\alpha_s$ 
is tedious since many Feynman diagrams contribute. 
Fig. \ref{twodiagrams} illustrates two such diagrams out of 62.
 The amplitude develops both a real and an imaginary part as in the well known DVCS case. 
No divergence nor end-point singularity plagues the validity of our approach in an obvious way. This fact 
can be confronted with the somehow similar process of the diffractive pion dissociation into two jets considered in
Ref.~\cite{BISS} for which the scattering amplitude already 
at the Born level suffers from an end-point singularities.

\begin{figure}[!h]
\vspace*{1cm}
\includegraphics[width=8cm]{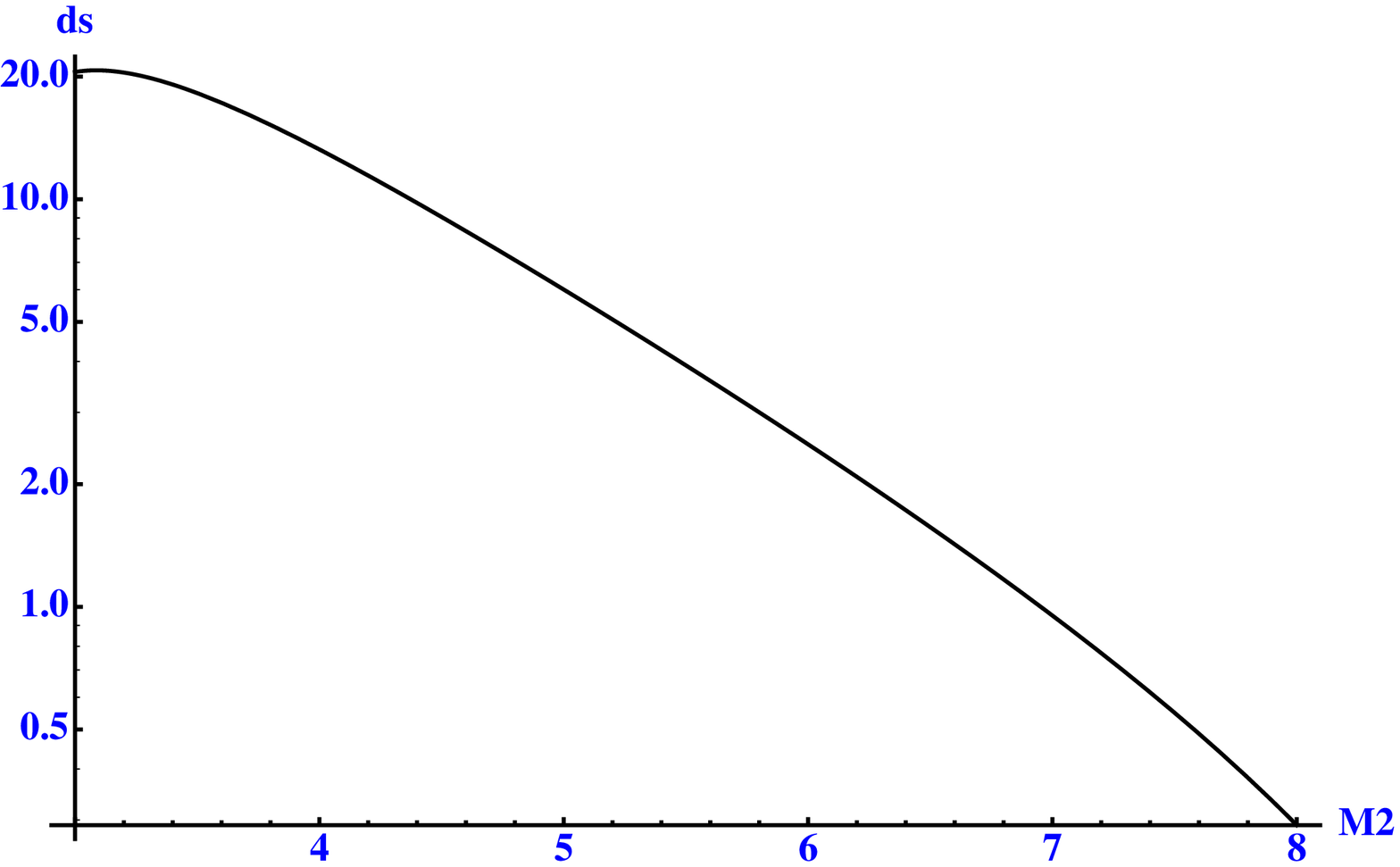}
\includegraphics[width=8cm]{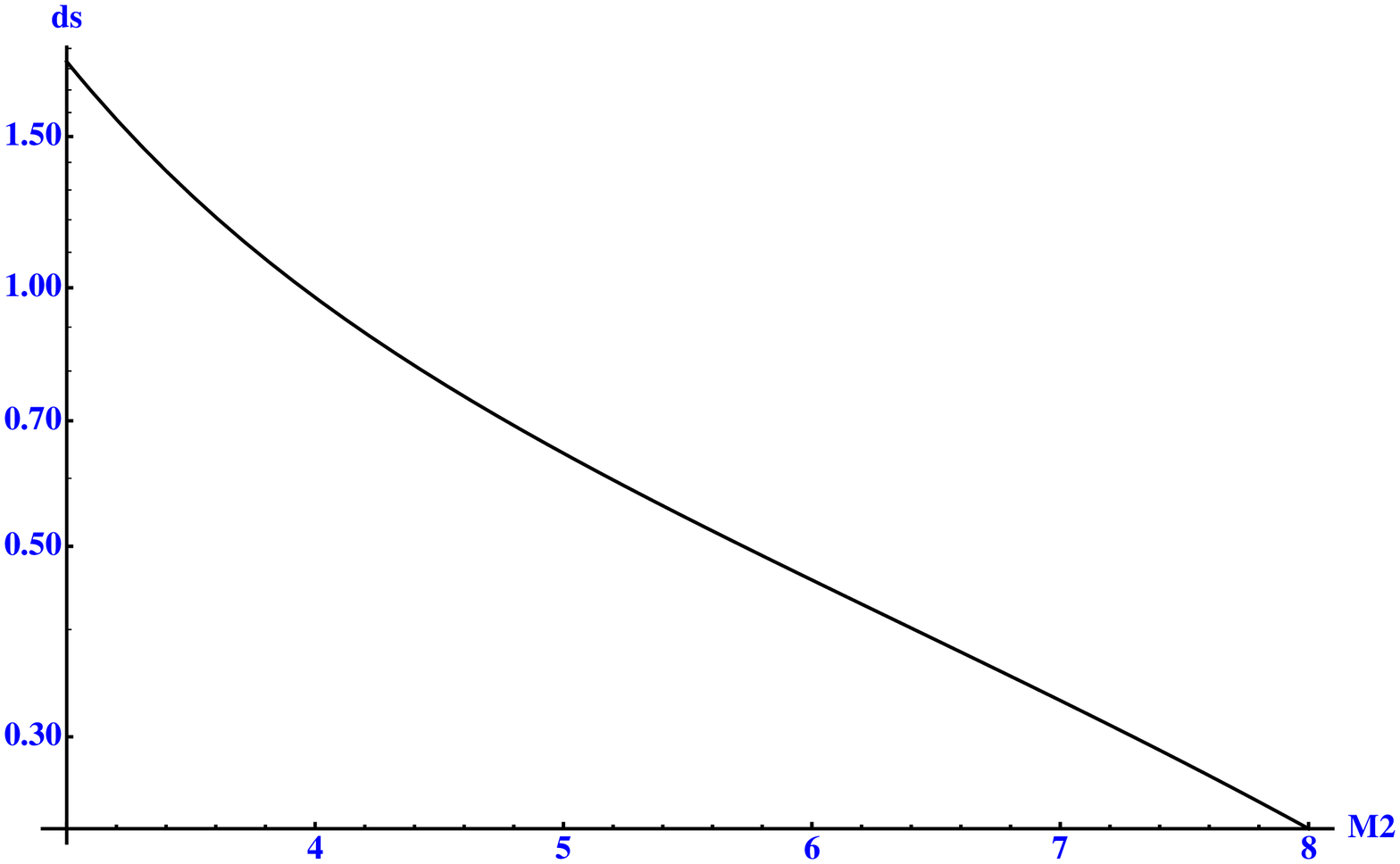}
\vspace{.3cm}
\caption{$M^2_{\pi\rho}$ dependence of the differential cross section 
in (nb.GeV$^{-2}$) at $S_{\gamma N}$ = 20 GeV$^2$ (left figure) and 
at $S_{\gamma N}$ = 100 GeV$^2$ (right figure).}
\label{result20}
\end{figure}

\begin{figure}[!h]
\begin{center}
\vspace{.5cm}
\includegraphics[width=8cm]{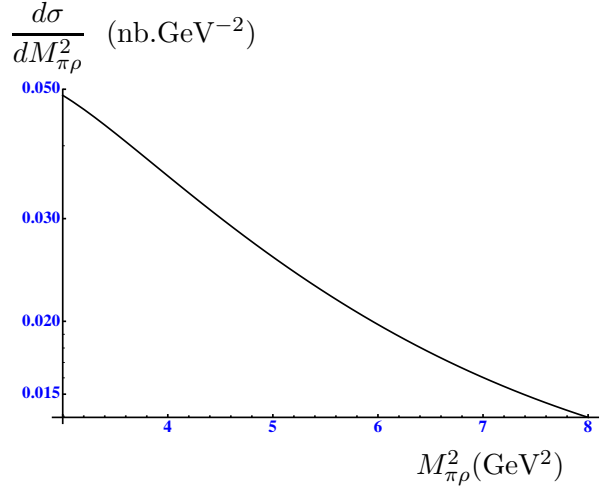}
\vspace{.3cm}
\caption{$M^2_{\pi\rho}$ dependence of the differential cross section 
in (nb.GeV$^{-2}$) at $S_{\gamma N}$ = 200 GeV$^2$.}
\label{result200}
\end{center}
\end{figure}

We performed a fairly detailed phenomenological study to judge the feasibility of the measurement. An example of the result are predictions for the integrated cross sections shown on Figs. \ref{result20}, \ref{result200}.
From our results, we conclude that the experimental search is promissing, both at low real or almost real photon energies within the JLab@12GeV upgraded facility, with the nominal  effective luminosity generally expected ($\mathcal{L} \sim 10^{35}$ cm$^2$.s$^{-1}$) and at higher photon energies with the  Compass experiment at CERN. These two energy ranges should give complementary information on the chiral-odd GPD $H_T(x,\xi,t)$. Namely, the large $\xi$ region may be scrutinized at JLab and the smaller $\xi$ region may be studied at COMPASS.
here also transversity GPDs can be measured in an unpolarized target experiment.
\psfrag{ds}{\raisebox{.5cm}{{\hspace{-.6cm}$\displaystyle \frac{d \sigma}{d M_{\pi \rho}^2}$ \hspace{0cm}{ (nb.GeV$^{-2}$)}}}}
\psfrag{M2}{\raisebox{-.7cm}{{\hspace{-2.5cm}$M_{\pi \rho}^2$(GeV$^2$)}}}

 \ack
We are grateful to D.Yu Ivanov, R. Enberg and O.V. Teryaev for their contributions to the results presented in section 3.1. 
This work is partly supported by the Polish Grant N202 249235.


\section*{References}
\begin{thebibliography}{99}

\bibitem{tra}
  J.~P.~Ralston and D.~E.~Soper,
  Nucl.\ Phys.\ B {\bf 152} (1979) 109;
  X.~Artru and M.~Mekhfi,
  Z.\ Phys.\ C {\bf 45} (1990) 669;
  J.~L.~Cortes, B.~Pire and J.~P.~Ralston,
  Z.\ Phys.\ C {\bf 55} (1992) 409;
  R.~L.~Jaffe and X.~D.~Ji,
  Phys.\ Rev.\ Lett.\  {\bf 67} (1991) 552.

\bibitem{Barone} 
  V.~Barone, A.~Drago and P.~G.~Ratcliffe,
  Phys.\ Rept.\  {\bf 359} 1 (2002) 1;
  M.~Anselmino,
  arXiv:hep-ph/0512140.
  
  \bibitem{Braun}
 B.~L.~Ioffe and A.~V.~Smilga,
  Nucl.\ Phys.\  B {\bf 232}, 109 (1984);
I.~I.~Balitsky, V.~M.~Braun and A.~V.~Kolesnichenko,
Nucl.\ Phys.\ B {\bf 312}, 509 (1989);
 P.~Ball, V.~M.~Braun and N.~Kivel,
  Nucl.\ Phys.\  B {\bf 649}, 263 (2003).


 \bibitem{PS}  
   B.~Pire and L.~Szymanowski,
  Phys.\ Rev.\ Lett.\  {\bf 103} (2009) 072002.


  
\bibitem{defDiehl}
  M.~Diehl,
  Eur.\ Phys.\ J.\ C {\bf 19} (2001) 485.

\bibitem{DGP}  
M.~Diehl, T.~Gousset and B.~Pire,
  Phys.\ Rev.\  D {\bf 59} (1999) 034023;
  J.~C.~Collins and M.~Diehl,
  Phys.\ Rev.\  D {\bf 61}  (2000) 114015.
  
  \bibitem{liuti}
   S.~Ahmad, G.~R.~Goldstein and S.~Liuti,
  Phys.\ Rev.\  D {\bf 79}  (2009) 054014;
S.~V.~Goloskokov and P.~Kroll,
Eur.\ Phys.\ J.\  C {\bf 65} (2010) 137.
  
\bibitem{IPST}   
D.~Y.~Ivanov, B.~Pire, L.~Szymanowski and O.~V.~Teryaev,
  Phys.\ Lett.\  B {\bf 550} (2002) 65.

\bibitem{eps}
 R.~Enberg, B.~Pire and L.~Szymanowski,
  Eur.\ Phys.\ J.\  C {\bf 47} (2006) 87.
  
\bibitem{GG} 
 L.~P.~Gamberg and G.~R.~Goldstein,
  Phys.\ Rev.\ Lett.\  {\bf 87} (2001) 242001
  [arXiv:hep-ph/0107176].
  
  \bibitem{EPSSW} 
   M.~El.~Beiyad, B.~Pire, M.~Segond, L.~Szymanowski and S.~Wallon,
  Phys.\ Lett.\  B {\bf 688}, 154 (2010)
and
arXiv:0911.2611 [hep-ph].


 \bibitem{kumano}
   S.~Kumano, M.~Strikman and K.~Sudoh,
  Phys.\ Rev.\  D {\bf 80} (2009) 074003.
  
\bibitem{Anselmino}
  M.~Anselmino \textit{et al.},
  Phys.\ Rev.\ D {\bf 75} (2007) 054032.

   
  

\bibitem{Pasq}
 S.~Scopetta,
  Phys.\ Rev.\  D {\bf 72} (2005) 117502;
  M.~Pincetti, B.~Pasquini and S.~Boffi,
  Phys.\ Rev.\  D {\bf 72} (2005) 094029 and Czech.\ J.\ Phys.\  {\bf 56} (2006) F229;
  M.~Wakamatsu,
  Phys.\ Rev.\  D {\bf 79} (2009) 014033;
  D.~Chakrabarti, R.~Manohar and A.~Mukherjee,
  Phys.\ Rev.\  D {\bf 79} (2009) 034006.
   M.~Gockeler {\it et al.}  [QCDSF Collaboration and UKQCD Collaboration],
   Phys.\ Rev.\ Lett.\  {\bf 98} (2007) 222001 and
  Phys.\ Lett.\  B {\bf 627} (2005) 113.

\bibitem{impact}
   M.~Burkardt,
  Phys.\ Rev.\  D {\bf 62} (2000) 071503;
 J.~P.~Ralston and B.~Pire,
  Phys.\ Rev.\  D {\bf 66} (2002) 111501;
  M.~Diehl,
  Eur.\ Phys.\ J.\  C {\bf 25} (2002) 223
  [Erratum-ibid.\  C {\bf 31} (2003) 277];
   M.~Burkardt,
  Phys.\ Rev.\  D {\bf 72} (2005) 094020;
   M.~Diehl and Ph.~Hagler,
  Eur.\ Phys.\ J.\  C {\bf 44} (2005) 87;
   A.~Mukherjee, D.~Chakrabarti and R.~Manohar,
  AIP Conf.\ Proc.\  {\bf 1149} (2009) 533
  [arXiv:0902.1461 [hep-ph]].

\bibitem{BISS} V.~M.~Braun, D.~Y.~Ivanov, A.~Schafer and L.~Szymanowski,
  Phys.\ Lett.\  B {\bf 509} (2001) 43
  [arXiv:hep-ph/0103275];
V.~M.~Braun, D.~Y.~Ivanov, A.~Schafer and L.~Szymanowski,
  Nucl.\ Phys.\  B {\bf 638} (2002) 111
  [arXiv:hep-ph/0204191].



\end{thebibliography}
\end{document}